# Geometric spin echo under zero field

Yuhei Sekiguchi, Yusuke Komura, Shota Mishima, Touta Tanaka, Naeko Niikura
and Hideo Kosaka*
Yokohama National University, 79-5 Tokiwadai, Hodogaya, Yokohama 240-8501, Japan
*kosaka-hideo-yp@ynu.ac.jp

Spin echo is a fundamental tool for quantum registers and biomedical imaging. It is believed that a strong magnetic field is needed for the spin echo to provide long memory and high resolution since a degenerate spin cannot be controlled or addressed under a zero magnetic field. While a degenerate spin is never subject to dynamic control, it is still subject to geometric control. Here we show the spin echo of a degenerate spin subsystem, which is geometrically controlled via a mediating state split by the crystal field, in a nitrogen vacancy center in diamond. The demonstration reveals that the degenerate spin is protected by inherent symmetry breaking called zero-field splitting. The geometric spin echo under zero field provides an ideal way to maintain the coherence without any dynamics, thus opening the way to pseudo-static quantum random access memory and non-invasive biosensors.

Nitrogen-vacancy (NV) centers in diamond provide promising platforms for quantum related technologies. The coherence time of an electron spin in an NV center has been shown to reach millisecond-order even at room temperature[1-4]. Arbitrary state preparation, single-/two-qubit control, entanglement generation[5,6,7] and quantum teleportation[8] have also been achieved using the NV spin together with proximate nuclear spins. Quantum memory, however, must possess two contradictory qualities: noise resilience and controllability. Symmetric states are known to be fragile by themselves but become noise resilient with specific symmetry-breaking operations. Such symmetry breaking inherently occurs in an NV center, where the ground-state electrons form triplet states with spin-1 angular momentum. The zero-field splitting due to the spin-spin interaction breaks the symmetry to push only the magnetic quantum number $m_S = 0$ state far below the degenerate $m_S = \pm 1$ states[9]. The zero-field split state provides a dynamic path to manipulate the geometric phase of the logical qubit based on the $m_S = \pm 1$ states[10,11]. The same situation is seen in photon polarization[12,13], which is also spanned by degenerate $m_S = \pm 1$ states representing circular polarizations with spin-1 angular momentum. This correspondence allows spontaneous entanglement generation[14] and entangled absorption[15] between single photon and electrons in an NV center in diamond. The correspondence also allows a microwave to be applied to manipulate an electron spin $\pm 1$ subsystem with arbitrary polarizations[16]. Even when the subsystem is degenerate enough to avoid a field to drive the state, it is possible to geometrically control[17,18] the geometric phase or the Berry phase[19]. Geometric gate operation has been proposed for performing holonomic quantum computation with built-in noise resilience[20-23]. The geometric phase has been experimentally observed in molecular ensembles[24,25], in a single superconducting qubit[26], and in a single NV center in diamond[10,11]. It has been shown that the geometric phase gate offers fast and precise control over the geometric phase, disrupts environmental interaction with multiple pulse echo sequences[17], and even offers a universal set of quantum logic gates[10].

Although those demonstrations introduced an energy gap to the qubit for controllability, here we show that it is possible to control a degenerate logical qubit, which we call a geometric spin qubit, by a purely geometric gate operation and that it can be protected by zero-field splitting with the help of a time-reversal operation, which we call geometric spin echo.

## Results
**System and scheme**. Our experimental demonstration of geometric spin control is based on the application of resonant microwave to electron spin in a diamond NV center under a zero magnetic field. The electron spin system in an NV center is described by the following Hamiltonian.

$$H = DS_z^2 + E_x(S_x^2 - S_y^2) + E_y(S_xS_y + S_yS_x) + \gamma_e \mathbf{B}_{ext} \cdot \mathbf{S} + A_z S_z I_z \tag{1}$$

where $\mathbf{S} = (S_x, S_y, S_z)$ is the spin-1 operator of the vacancy electron spin, $I_z$ is the z component of the spin-1/2 operator of the $^{14}$N nuclear spin, $D = (2\pi \times)$ 2.87 GHz is the axial zero-field splitting, $E_x$ ($E_y$) is the $x$ ($y$) component of the transverse zero-field splitting caused by a crystal strain, $\gamma_e$ is the gyromagnetic ratio of the electron spin, $\mathbf{B}_{ext}$ is the external magnetic field and $A_z = (2\pi \times)$ 2.175 MHz is the z component of the hyperfine interaction between the electron spin and the $^{14}$N nuclear spin. The $x$, $y$ components of the hyperfine interaction and magnetic field, which contribute to the second-order perturbation, are negligible owing to the large zero-field splitting. Note that in this paper we omit the Planck constant $\hbar$ for simplicity.

If we take only the first term of equation (1) as a dominant term, the ground state forms triplet states consisting of degenerate $|m_S = \pm 1\rangle$ states, which serve as logical qubit basis states, and a zero-field split $|m_S = 0\rangle$ state, which serves as an ancillary state for the geometric operation as shown in Fig. 1a. Based on the Jaynes-Cummings model, the interaction Hamiltonian with a microwave resonant to the energy gap between the $|\pm 1\rangle$ and $|0\rangle$ states is described as $H_{MW}(t) = \frac{\Omega(t)}{2} S_x$, where $\Omega(t)$ denotes the Rabi frequency. We define the polarization of the microwave as a linear polarization oriented toward $+x$. The spin 1 operator $S_x = |0\rangle\langle B| + |B\rangle\langle 0|$ indicates the state exchange between the bright state $|B\rangle = \frac{1}{\sqrt{2}}(|+1\rangle + |-1\rangle)$ and $|0\rangle$, while the dark state $|D\rangle = \frac{1}{\sqrt{2}}(|+1\rangle - |-1\rangle)$ remains unchanged (Fig. 1b). After a round trip time $T$, defined as $\int_0^T \Omega(t)dt = 2\pi$, the bright state $|B\rangle$ evolves as $\exp(-i\int_0^T H_{MW}(t)dt)|B\rangle = \exp(-i\pi\sigma_x^{\{B,0\}})|B\rangle = -|B\rangle$, where $\sigma_x^{\{B,0\}} = |0\rangle\langle B| + |B\rangle\langle 0|$ denotes the Pauli operator in the $|B\rangle$-$|D\rangle$ subspace. Note that the prefactor $-1$ is nothing but a global phase in the $|B\rangle$-$|0\rangle$ subspace, whereas in the $|\pm 1\rangle$ subspace the global phase serves as a relative phase called the geometric phase for a geometric spin qubit. This geometric operation is represented as a $\pi$ rotation around the $|B\rangle$-$|D\rangle$ axis or the $x$ axis in the $|\pm 1\rangle$ subspace as $\exp\left(-i\frac{\pi}{2}\sigma_x^{\{\pm 1\}}\right) = -i\sigma_x^{\{\pm 1\}}$ as shown in Fig. 1c. The pulse sequences used in the demonstrations are summarized in Fig. 1d (Methods).

**Rabi oscillation and Ramsey interference**. A series of experiments for calibrating the condition to achieve the geometric spin echo are performed under a zero magnetic field. The Rabi oscillation experiment determines the $\pi$ pulse width required to flip the spin states between the $|B\rangle$ state and the $|0\rangle$ state (Fig. 2a). The oscillation conforms to the theory considering hyperfine coupling between the electron spin at the vacancy and the nuclear spin at the nitrogen ($^{14}$N) that comprises the NV. The Ramsey interference experiment or the free-induction decay indicates that the hyperfine coupling induces electron spin precession to alter the $|B\rangle$ and $|D\rangle$ states at a frequency corresponding to twice the hyperfine coupling (Fig. 2b). The Gaussian decay of the envelope indicates the geometric spin coherence time $T_2^*$ to be 0.61 μs. The origin of the decoherence would be the coupling of the electron spin to a spin bath consisting of the proximate nuclear spins of the $^{13}$C isotopes.

**Geometric spin echo**. The disappeared Ramsey interference signal seen in Fig. 2b recovers as a geometric spin echo by the insertion of $2\pi$ pulse after 35 μs of time evolution and the signal reaches a maximum when the second evolution time equals the first (Fig. 3a). The result indicates that the geometric spin echo rephases the geometric spin as the conventional Hahn echo does the dynamic spin, even under complete degeneracy of the qubit space. Figure 3b shows the signal decay of the geometric spin echo under external magnetic fields of 0 mT (red squares), 0.04 mT (green circles), and 0.12 mT (blue triangles) measured along the NV axis. The coherence time is extended by the echo process to $T_2 = 83$ μs under a zero magnetic field, which is about 140 times longer than the $T_2^*$ of 0.61 μs. The echo coherence time drastically increases as decreasing the magnetic field from 0.12 mT, within which both of the Zeeman-split electron spin states are equally driven by the microwave. The $T_2$ in Fig. 3b is determined only by fitting with $\exp[-(t/T_2)^3]$ under the assumption that the population decay behaves as $\exp(-t/T_1)$, where $T_1 = 700$ μs is dominated by green laser leakage (Supplementary Fig. 1).

The echo coherence time $T_2$ as a function of an external magnetic field measured along the NV axis (Fig. 3c) agrees relatively well with theory based on the disjoint cluster method[27], which neglects electron spin flip owing to the large axial zero-field splitting[28], for a spin bath consisting of $^{13}$C isotopes with a natural abundance of 1.1% (Methods). Single and dimer $^{13}$C nuclear spins (nearest-neighbor nuclear spin pair) were taken into account as bath spins but the interaction between bath spins were neglected[29].

Figure 4a decomposes contributions of single and dimer nuclear spins to the decoherence. Note that the single nuclear spins dominate decoherence under relatively high magnetic field, while the dimer nuclear spins dominate under a zero magnetic field. In the case of single nuclear spins, the quantization axis defined by the electron spin hyperfine field is deviated by the external magnetic field. In the other case of dimer nuclear spins, the quantization axis is deviate by the dipolar magnetic field within the dimer (Fig. 4b). In any case, the electron spin inversion in the logical qubit space cannot completely time reverse the nuclear spin dynamics during the geometric spin echo. In other words, the time evolution operators with orthogonal electron spin states become incompatible or irreversible by the additional field as in equation (9).

**Discussions.** The deviation from the theory around 0 and 0.075 mT in Fig. 3c, on the other hand, is explained by the strain, or the transverse zero-field splitting, which couples the otherwise degenerate $m_S = \pm 1$ states to lift the degeneracy and further decouples the spin bath from the electron spin. The magnetic field to give maximum deviation 0.075 mT corresponds to the nitrogen hyperfine splitting 2.18 MHz. The range of the deviation ~0.02 mT is explained by considering the strain splitting 0.23 MHz and inhomogeneous broadening 0.43 MHz of the optically detected magnetic resonance (ODMR) spectrum (Supplementary Fig. 3a). The calculated magnetic-field dependence of the echo $T_2$ with the strain correction (Supplementary Fig. 3b) agrees well with the experiments, indicating that the geometric spin qubit is protected against decoherence not only by the axial zero-field splitting but also by the transverse zero-field splitting. In contrast that the axial zero-field splitting suppresses the transverse hyperfine interaction, which causes a bit-flip error $\sigma_x^{\{\pm 1\}}$ to the second-order perturbation, the transverse zero-field splitting suppresses the axial hyperfine interaction, which causes a phase-flip error $\sigma_z^{\{\pm 1\}}$, to the second-order perturbation[30].

Although the achieved geometric echo coherence time $T_2$ of 83 μs at the degeneracy is 140 times longer than the free induction decay of 0.61 μs, it is several times shorter than the measured population decay time $T_1$ of 700 μs for the geometric spin state relaxing from $|B\rangle$ to $|0\rangle$ (Supplementary Fig. 1). The discrepancy between $T_1$ and $T_2$ can be compensated by excluding the effect of dimers not only by decreasing the abundances of $^{13}$C isotopes but also by suppressing dimers into their singlet state to be spin transparent. Since the common dynamical qubit defined in the 0/-1 subspace uses only one component of the +1/-1 subspaces, which couples with the $^{13}$C spin bath, the Ramsey coherence time $T_2$* should be double of the geometric qubit. However, the geometric qubit should surpasses the dynamical qubit with the multi-pulse echo[17,31] even in the absence of magnetic field, as previously demonstrated in the presence of magnetic field[32]. Since the dynamical Hahn echo cannot inverse the +1/-1 subspace, it cannot reverse time to recover the original state even in the absence of magnetic field. In contrast, the developed geometrical qubit defined in the +1/-1 subspace is space-inversed by the geometrical echo and thus time-reversed to decouple the spin bath.

We demonstrated the geometric spin echo of a degenerate geometric spin qubit via the ancillary state in a diamond nitrogen vacancy center. The geometric spin echo recovered the coherence imprinted in the degenerate subspace after 140 times the free induction decay time. The theoretical analysis indicates that the geometric spin qubit is three-dimensionally protected against decoherence by the axial and transverse zero-field splittings with the help of time reversal, leaving decoherence due to dimer $^{13}$C nuclear spins under a zero magnetic field. The purely geometric spin qubit is not only robust against noise caused by the spin bath but also robust against control error, and thus is suitable for a memory qubit used in quantum information and quantum sensing of magnetic, electric or strain fields for bio-medical imaging.

**Methods**
**Experimental setup**. We used for the experiments a native NV center in type-IIa high-pressure-high-temperature (HPHT) grown bulk diamond with a ⟨001⟩ crystal orientation (from Sumitomo Electric) without any irradiation or annealing. This diamond has NV centers (~$10^{12}$ cm$^{-3}$) and nitrogen impurities called P1 centers (less than 1 ppm), leading to relatively higher density than a chemical-vapor-deposition (CVD) grown diamond. A negatively charged NV center located at about 30 μm below the surface was found using a confocal laser microscope. A 25-μm copper wire mechanically attached to the surface of the diamond was used to apply a microwave for the optically detected magnetic resonance (ODMR) measurement. An external magnetic field at an angle of 70° to the NV axis was applied to compensate the geomagnetic field of about 0.045 mT using a permanent magnet. Careful orientation of the magnet was conducted with monitoring of
the ODMR spectrum within 0.1 MHz. The Rabi oscillation and Ramsey interference were also used to fine-tune the field. The NV center used in the experiment showed no hyperfine splitting caused by $^{13}$C nuclear spins exceeding 0.1 MHz. All experiments were performed at room temperature.

**Pulse sequences**. The pulse sequences used in the demonstrations are summarized in Fig. 1e. Green light irradiation for 3 μs initializes the electron system into the ancillary state $|0\rangle$, which is followed by microwave pulses resonant to the zero-field splitting $D$ with pulse patterns depending on the experiment. The Rabi oscillation between the bright state $|B\rangle$ and the ancillary state $|0\rangle$ was first observed to determine the π pulse width. The Ramsey interference was then observed to confirm the coherence between logical qubit basis states $|+1\rangle$ and $|-1\rangle$ by letting those superposition states revolve between $|B\rangle$ and $|D\rangle$ during the time between the two π pulses, instead of two π/2 pulses, as is used for the conventional Ramsey interference. Finally, we demonstrated the geometric spin echo by applying a 2π pulse in the middle of the free precession, instead of a π pulse as is used for the conventional Hahn echo. Despite the differences in the scheme, the random phase shift caused by the spin bath rephased back into the initial state as is schematically shown in the inset of Fig. 1e. The photon counts during the first 300 ns normalized by those during the last 2 μs of the 3-μs green laser irradiation for the next initialization were used to measure the population in the $|0\rangle$ state. All the pulse sequences and photon counts were managed by an FPGA-based control system developed by NEC communications.

**Calculation model**. For the calculation of the magnetic field dependence of $T_2$ echo time under a zero magnetic field, we neglected the $T_1$ relaxation time since it is sufficiently longer than $T_2$. The decoherence is therefore dominated by the effects of single $^{13}$C nuclear spins and dimers (nearest-neighbor nuclear spin pair). Because of the large discrepancy in the Zeeman energy between the electron spin and the bath spins, we neglected the electron spin bit flip induced by the bath spins and attributed the coherence of the electron spin to that of the bath spins[27,29],

$$S(2\tau) = \mathrm{Tr}\big[U^{(-)}(\tau)U^{(+)}(\tau)\rho_N U^{(-)\dagger}(\tau)U^{(+)\dagger}(\tau)\big], \tag{2}$$

where $U^{(\pm)}$ is the time evolution operator depending the electron spin state $|\pm\rangle$ and $\rho_N$ is the spin bath state operator; at the high-temperature limit, $\rho_N = \frac{\mathbb{1}^{\otimes N}}{2^N}$ ($N$: the number of bath spins). We considered only single and dimer $^{13}$C nuclear spins as bath spins and neglected the interaction between bath spins[28]. The electron spin echo signal could thus be factorized into individual bath spins as follows:

$$S(2\tau) \approx \prod_j S_{\mathrm{single},j}(2\tau) \prod_k S_{\mathrm{dimer},k}(2\tau), \tag{3}$$

where $S_{\mathrm{single},j}(2\tau)$ is the $j$th single $^{13}$C nuclear spin contribution and $S_{\mathrm{dimer},k}(2\tau)$ is the $k$th dimer $^{13}$C nuclear spin contribution.

**Electron spin decoherence**. The dipole-diploe interaction between $i$ spin and $j$ spin is defined as

$$H_{\text{dip},ij} = \mathbf{S}_i \cdot \mathbf{A}_{ij} \cdot \mathbf{S}_j$$
$$= \frac{\mu_0}{4\pi} \frac{\gamma_i \gamma_j}{r_{ij}^3} \left( \mathbf{S}_i \cdot \mathbf{S}_j - 3 \frac{(\mathbf{S}_i \cdot \mathbf{r}_{ij})(\mathbf{r}_{ij} \cdot \mathbf{S}_j)}{r_{ij}^2} \right), \quad (4)$$

where $\mu_0$ is the vacuum permeability, $\gamma_i$ ($\gamma_j$) is the gyromagnetic ratio of the $i$ ($j$) spin (electron spin: $\gamma_e = -1.76 \times 10^{11}$ rad s$^{-1}$ T$^{-1}$, $^{13}$C nuclear spin: $\gamma_c = 6.73 \times 10^7$ rad s$^{-1}$ T$^{-1}$), $\mathbf{S}_i$ ($\mathbf{S}_j$) is the spin operator of the $i$ ($j$) spin, and $\mathbf{r}_{ij}$ is the displacement of the $i$ spin from the $j$ spin. In the following, we describe $\mathbf{S}$ ($\mathbf{I}_j$) as a spin-1 operator of the electron spin (spin-1/2 operator of $j$th $^{13}$C nuclear spin). Hyperfine interaction conditioned by the electron spin eigenstates $|n = 0, \pm\rangle$ of the Hamiltonian in equation (1) can be represented as

$$H_{\text{dip},ej} \approx \sum_n |n\rangle\langle n| \otimes \mathbf{A}_{ej}^{(n)} \cdot \mathbf{I}_j, \quad (5)$$

where $\mathbf{A}_{ej}^{(n)} \equiv \langle n|\mathbf{S}|n\rangle \cdot \mathbf{A}_{ej}$ is the electron spin hyperfine field depending on the electron spin state $|n\rangle$. Then, we can define the effective magnetic field, Hamiltonian and time evolution operator depending on the electron spin state $|n\rangle$ as

$$\mathbf{B}_j^{(n)} = \mathbf{B}_{\text{ext}} + \frac{\mathbf{A}_{ej}^{(n)}}{\gamma_c}, \quad (6)$$

$$H_{\text{single},j}^{(n)} = \gamma_c \mathbf{B}_j^{(n)} \cdot \mathbf{I}_j, \quad (7)$$

$$U_{\text{single},j}^{(n)}(\tau) = e^{-iH_j^{(n)}\tau}. \quad (8)$$

The echo signal given by the single $j$th $^{13}$C spin is[33]

$$S_{\text{single},j}(2\tau) = \frac{1}{2} \text{Tr}\left[ U_{\text{single},j}^{(-)}(\tau) U_{\text{single},j}^{(+)}(\tau) U_{\text{single},j}^{(-)\dagger}(\tau) U_{\text{single},j}^{(+)\dagger}(\tau) \right]$$
$$= 1 - \frac{\left|\mathbf{B}_j^{(+)} \times \mathbf{B}_j^{(-)}\right|^2}{\left|\mathbf{B}_j^{(+)}\right|^2 \left|\mathbf{B}_j^{(-)}\right|^2} \sin\left( \frac{\gamma_c \left|\mathbf{B}_j^{(+)}\right|^2 \tau}{2} \right) \sin\left( \frac{\gamma_c \left|\mathbf{B}_j^{(-)}\right|^2 \tau}{2} \right). \quad (9)$$

These equations indicate that a single nuclear spin does not decohere the electron spin under a zero magnetic field after the spin echo since the time evolution operators depending on the electron spin state are compatible, while it does decohere the electron spin under a non-zero magnetic field. On the other hand, the effective magnetic field, Hamiltonian and time evolution operator of the $k$th dimer are

$$H_{\text{dimer},k}^{(n)} = \gamma_c \mathbf{B}_k^{(n)} \cdot \mathbf{I}_{k0} + \gamma_c \mathbf{B}_k^{(n)} \cdot \mathbf{I}_{k1} + \mathbf{I}_{k0} \cdot \mathbf{A}_{k0k1} \cdot \mathbf{I}_{k1}, \quad (10)$$

$$U_{\text{dimer},k}^{(n)}(\tau) = e^{-iH_{\text{dimer},k}^{(n)}\tau}, \quad (11)$$

where $\mathbf{I}_{k0}$, $\mathbf{I}_{k1}$ are 0th, 1st $^{13}$C spins consisting the dimer and we suppose that the effective magnetic field is applied equally to each nuclear spin. The echo signal given by the $k$th dimer is

$$S_{\text{dimer},k}(2\tau) = \frac{1}{4}\text{Tr}\left[U^{(-)}_{\text{dimer},k}(\tau)U^{(+)}_{\text{dimer},k}(\tau)U^{(-)\dagger}_{\text{dimer},k}(\tau)U^{(+)\dagger}_{\text{dimer},k}(\tau)\right]. \tag{11}$$

The spin bath configuration is generated by randomly placing $^{13}$C isotopes with natural abundance of 1.1% at a distance within 4 nm of the vacancy electron (Supplementary Fig. 2). $T_2$ is determined by fitting with $\exp[-(t/T_2)^3]$.

**Transverse zero-field splitting**. If we neglect the transverse magnetic field for simplicity, the Hamiltonian in equation (1) can be rewritten as $H = E_x \sigma_x^{\{\pm 1\}} + E_y \sigma_y^{\{\pm 1\}} + \left(\gamma_e B_{\text{ext},z} + A_z \langle m_I | I_z | m_I \rangle\right)\sigma_z^{\{\pm 1\}}$ thus generating the following eigenenergies depending on the nitrogen nuclear spin states $m_I = 0, \pm 1$

$$\varepsilon^{m_I}_\pm = \pm \sqrt{E_x^2 + E_y^2 + \left(\gamma_e B_{\text{ext},z} + A_z \langle m_I | I_z | m_I \rangle\right)^2}, \tag{12}$$

$$\varepsilon_0 = 0, \tag{13}$$

and eigenstates

$$|+\rangle^{m_I} = \cos\frac{\theta^{m_I}}{2}|+1\rangle + e^{i\phi^{m_I}}\sin\frac{\theta^{m_I}}{2}|-1\rangle, \tag{14}$$

$$|-\rangle^{m_I} = \sin\frac{\theta^{m_I}}{2}|+1\rangle + e^{-i\phi^{m_I}}\cos\frac{\theta^{m_I}}{2}|-1\rangle, \tag{15}$$

$$|0\rangle, \tag{16}$$

where $\theta^{m_I}$, $\phi^{m_I}$ are the polar and azimuth angles of the Bloch sphere spanned by $|m_S = \pm 1\rangle$,

$$\theta^{m_I} = \arccos\left(\frac{\gamma_e B_{\text{ext},z} + A_z \langle m_I | I_z | m_I \rangle}{\sqrt{E_x^2 + E_y^2 + \left(\gamma_e B_{\text{ext},z} + A_z \langle m_I | I_z | m_I \rangle\right)^2}}\right), \tag{17}$$

$$\phi^{m_I} = \arccos\left(\frac{E_x}{\sqrt{E_y^2 + E_y^2}}\right). \tag{18}$$

As the $z$ component of the energy splitting $\gamma_e B_{\text{ext},z} + A_z \langle m_I | I_z | m_I \rangle$ decreases compared to the transverse zero-field splitting $\sqrt{E_x^2 + E_y^2}$, where $E_x$ ($E_y$) is $x$ ($y$) component of the energy splitting, the $m_S = \pm 1$ states couple strongly and finally become completely coupled states $|\pm\rangle = \left(|+1\rangle + e^{\pm i\phi}|-1\rangle\right)/\sqrt{2}$. The transverse zero-field splitting under the condition thus decouples the interaction between the electron spin and bath spins to suppress the decoherence effect by decreasing the electron spin hyperfine field $\mathbf{A}^{(n)}_{ej} \equiv \langle \pm | \mathbf{S} | \pm \rangle \cdot \mathbf{A}_{ej}$. This protection effect explains the enhancement seen in Fig. 3b around 0 and 0.075 mT, where the nitrogen hyperfine field cancels out the $z$ component of the external magnetic field on the electron spin. The enhancement becomes prominent by initializing the nitrogen nuclear spin state to $m_I = 0$ state under a zero magnetic field, at which point the electron spin echo coherence time $T_2$ drastically increase.

**Acknowledgements**

We thank Yuichiro Matsuzaki, Fedor Jelezko, Burkhard Scharfenberger, Kae Nemoto, William Munro, Norikazu Mizuochi, and Joerg Wrachtrup for their discussions and experimental help. This work was supported by the NICT Quantum Repeater Project, by the FIRST Quantum Information Project, and by a Grant-in-Aid for Scientific Research (A)-JSPS (No. 24244044).

**Author contributions**

The experiment was designed and analysed by Y. S., Y. K. and H. K. Measurements were made by Y. S. and Y. K. S. M. and T. T. and N. N. supported experiments in technical matters. H. K. supervised experiments. Y. S., Y. K. and H. K. wrote the paper.

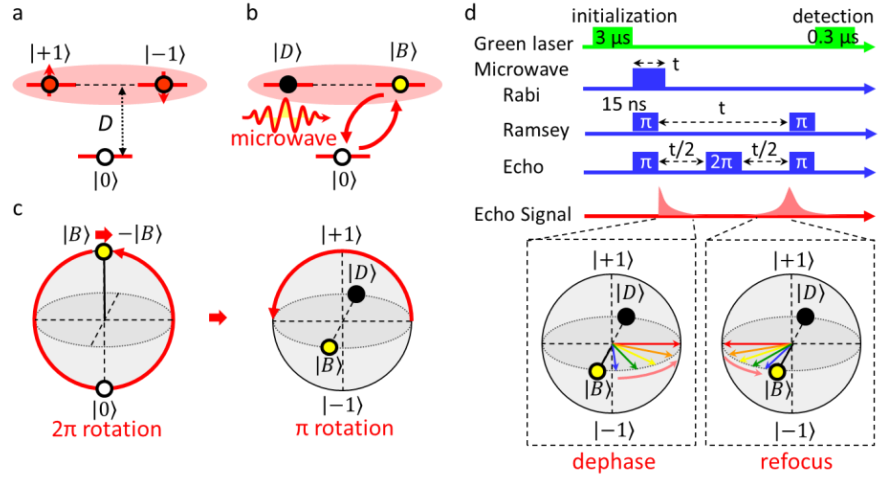

**Figure 1 | Geometric operation and experimental sequence.** (**a**) Energy level diagram of an electron spin in an NV center under a zero magnetic field on the computational bases. (**b**) The logical bases $|\pm 1\rangle$ are transformed into bright $|B\rangle$ and dark $|D\rangle$ states, as defined by the microwave polarization. (**c**) Bloch sphere representation of a $2\pi$ rotation starting from the bright state $|B\rangle$ through the ancillary state $|0\rangle$ returning to $|B\rangle$ with an additional geometric phase factor -1. The geometric phase contributes to a $\pi$ rotation around the $|B\rangle$-$|D\rangle$ axis in the logical qubit space. (**d**) Pulse sequence used for the Rabi oscillation, Ramsey interference, and echo recovery. Insets illustrate the logical qubit dephasing and refocusing after the $2\pi$ pulse.

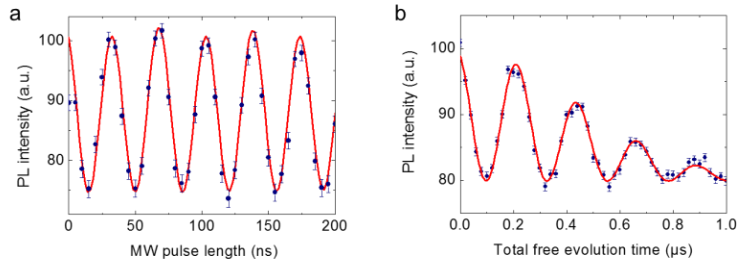

**Figure 2 | Rabi oscillation and Ramsey interference.** (**a**) The Rabi oscillation between the bright $|B\rangle$ and ancilla $|0\rangle$ states. The solid line shows the best theoretical fit to the data considering the hyperfine interaction with the nuclear spin of the nitrogen ($^{14}$N) comprising the NV. (**b**) The Ramsey interference between $|B\rangle$ and $|D\rangle$ states. The solid line shows the best theoretical fit with a Gaussian envelope. The precession frequency of 4.35MHz corresponds well to hyperfine splitting. Error bars are defined as the s.d. of the photon shot noise.

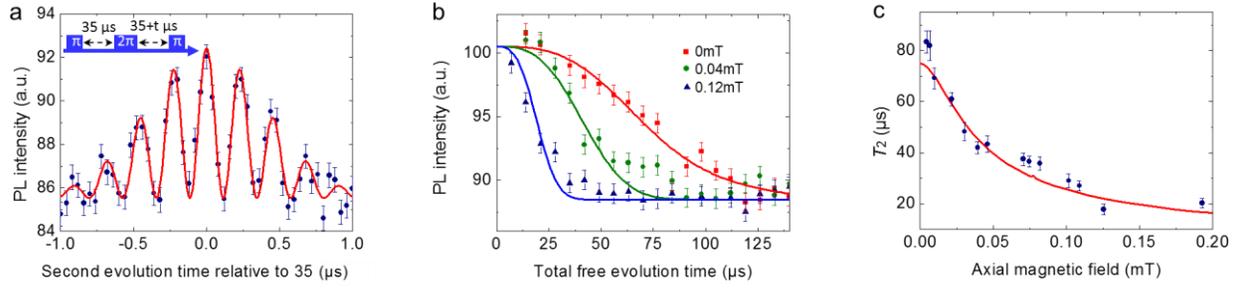

**Figure 3 | Geometric spin echo.** (**a**) Refocusing of the geometric echo signal by fixing the first evolution time at 35 μs and sweeping the second evolution time around 35 μs. The solid line shows the best theoretical fit with a Gaussian envelope. (**b**) Echo decays of the electron spin under magnetic fields of 0 mT (red squares), 0.04 mT (green circles) and 0.12 mT (blue triangles) measured along the NV axis. Solid lines show the best fitting curves $\exp[-(t/T_2)^3]$ integrated with $T_1$ decay. (**c**) Echo coherence decay time $T_2$ as a function of the magnetic field. The solid line shows the theoretical fits to the data (Methods). Error bars in (**a**), (**b**) are defined as the s.d. of the photon shot noise. Error bars in (**c**) are given by least squares fitting.

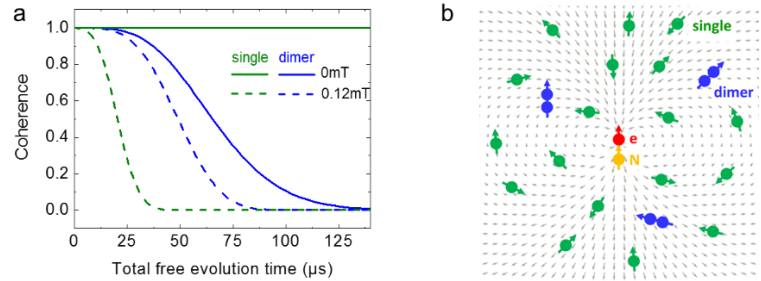

**Figure 4 | Single and dimer nuclear spins.** (**a**) Calculated geometric spin echo decay induced by bath spins decomposed into single (green) and dimer (blue) $^{13}$C nuclear spin contributions. Solid (dashed) lines shows the decay under a magnetic field of 0 mT (0.4 mT) measured along the NV axis. (**b**) Schematic spin bath configuration around the vacancy electron spin under a zero magnetic field. Single $^{13}$C nuclear spins are quantized along the electron spin hyperfine field (gray arrows). Dimer $^{13}$C nuclear spins are quantized along the dipolar magnetic field within each of the dimers.

# Supplementary Information

## Supplementary Figures

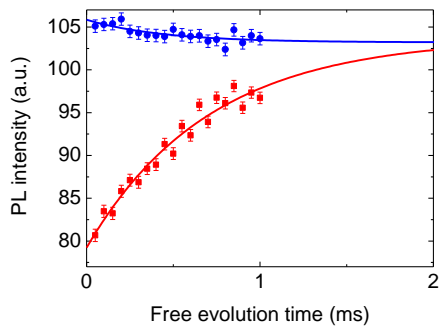

**Supplementary Figure 1 | The electron spin relaxation.** Time dependence of the pulsed ODMR signal prepared in the bright state $|B\rangle$ (red sqares) and the ancillary state $|0\rangle$ (blue circles). Solid lines show exponential decay fittings. The decay rate corresponded to the population decay time $T_1$ of 700 μs used in the fitting in Fig. 3c. Error bars are defined as the s.d. of the photon shot noise.

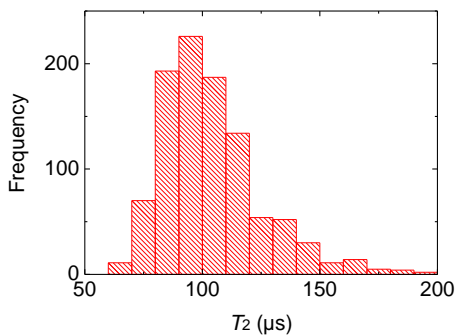

**Supplementary Figure 2 | Histogram of dimer $T_2$ under a zero magnetic field.** Theoretical distribution of the $T_2$ time under random dimer $^{13}$C nuclear spin bath configuration, where we only consider the dimers within 4 nm of electron spin. The bath configuration giving $T_2 = 75$ μs is used for the calculation of Fig. 3c and 4a.

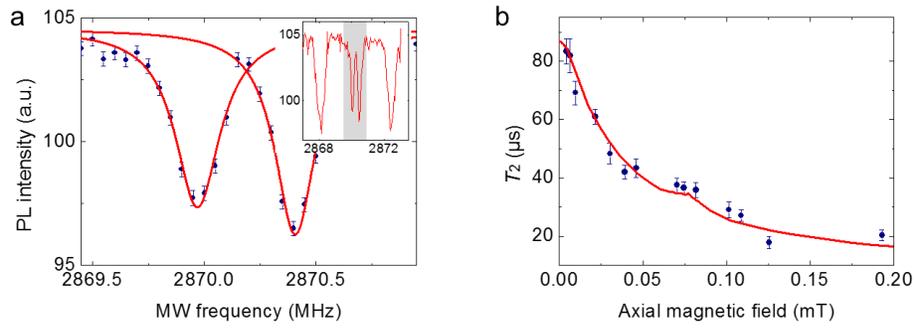

**Supplementary Figure 3 | Transverse zero-field splitting effect.** (**a**) Frequency dependence of the pulsed ODMR signal showing the strain splitting in the dip corresponding to the nuclear spin $m_\text{I} = 0$ state. Solid lines are Lorenzian curve fittings. Inset shows the full spectrum of the pulsed ODMR, where the center part (grey area) corresponds to the $m_\text{I} = 0$ state. (**b**) Echo coherence decay time $T_2$ as a function of magnetic field measured along the NV axis with considering the strain splitting of 0.23 MHz and inhomogeneous broadening of 0.43 MHz due to the hyperfine field from the spin bath. Error bars in (**a**) are defined as the s.d. of the photon shot noise. Error bars in (**b**) are defined as the s.d. of least squares fitting.